# Telecom data for efficient malaria interventions

Kristýna Tomšů
Real Impact Analytics
Brussels, Belgium
kristyna.tomsu@realimpactanalytics.com

Alexis Eggermont
Real Impact Analytics
Brussels, Belgium
alexis.eggermont@realimpactanalytics.com

Nicolas Snel
Real Impact Analytics
Brussels, Belgium
nicolas.snel@realimpactanalytics.com

## ABSTRACT
Telecom data is rich on mobility information and as such can be used to identify mobility patterns of people in near real time, enabling to build epidemiological models for understanding where epidemics might spread over time. Based on previous research, we have built an operational tool fed with telecom data which shows malaria risk flows in Zambia in near real time. It provides insights on which areas should eradicate malaria first in order to have a maximum impact on the overall country malaria flows, and it highlights regions that should coordinate their eradication efforts together. Such information is particularly relevant for countries like Zambia, which are getting close to malaria elimination and need to prevent its re-introduction into areas that are already malaria-free.

## 1. INTRODUCTION
Systematic efforts and innovative tools led to significant progress towards reduction in malaria burden, with a 47% decrease in malaria-attributable deaths since 2000 (WHO, 2014). However, as more countries are entering their pre-elimination and elimination phases, national malaria programs are facing new challenges specific to low-transmission settings. One of these is linked to human travel and its contribution to malaria transmission, as travelers from highly endemic regions can re-introduce the disease into areas already declared malaria-free or strengthen the transmission chains in areas with limited malaria exposure. Because of that, monitoring of human mobility is one of the key elements for National Elimination Malaria Strategies for countries in low-transmission settings.

However, little is known about current mobility patterns and reliable systematic insights regarding people's travel are not available for intervention planning. Currently, decision makers are relying on often inaccurate travel reports or outdated census data.

One possibility of tackling this problem is leveraging telecom data which is systematically collected, covers large proportions of population and is rich on information about individual and collective mobility. This data is routinely collected by mobile phone operators for billing purposes and, when aggregated to small geographical areas, can be merged with available malaria data. A substantial body of research shows that telecom data is a suitable data source for enriching malaria models.

Tatem et al. (2009) explored the contribution of visitors to mainland Tanzania to importation of malaria to Zanzibar. Wesolowski et al. (2012) used telecom data to estimate the number of infections exported and imported to individual settlements in Kenya, mapping the settlements into malaria sinks and sources. Tatem et al. (2014) used a simplified model to rank different areas by their contribution towards malaria risk export or import, with a focus on low-transmission settings in Namibia.

To build upon this research, we are developing an interactive user-friendly tool, following the tested methodology, but allowing for near real-time monitoring of population movement and related malaria risk flows as well as for prioritization of small-scale areas where malaria interventions would have the highest impact.

## 2. METHODOLOGY
Our methodology follows closely that of Wesolowski et al (2012) for data preparation and Tatem et al (2014) for relative risk estimation.

### 2.1. DATA
We are working with data from one of the leading operators in Zambia. The data used are the call detail records (CDRs), which the operator has to store for billing purposes. The CDRs contain a user id, time and location (at the level of transmitting cell tower) of every incoming or outgoing call, text or data usage.

For malaria, we use monthly case incidence data as reported through facilities and volunteering community healthcare workers. As the granularity of the telecom data is given by the density of the cell tower network, the incidence is calculated based on reported cases and population in the corresponding tower catchment areas, with population estimates extracted from the WorldPop project (http://www.worldpop.org.uk/).

### 2.2. DATA PROCESSING
The location granularity is given primarily by the cell tower distribution. However, since we are not interested in mobility within urban areas, towers within the same urban extent are merged into one settlement (for estimation of urban extents, average population density is used).

We estimate the home location of each user as the settlement at which they are most active during the weekends and at night. The home location is calculated over a period of 60 days. For identification of short-term trips, the home location is compared to the daily most used settlement. If a user is having their most used settlement away from the home location for a sequence of at least two days, we count it as a trip. As malaria spreads at nights, the length of a trip is estimated by the number of nights spent away from home. The individual trips are then aggregated for a total number of trips between each pair of settlements.

For each settlement A and a pair of settlements A and B, we estimate two types of flows: 1) returning residents – the people who have their home location in A, left for a trip to B and eventually returned to their home in A, and 2) visitors – the people who have their home location in B, they came for trip to A and eventually returned to B. Note that returning visitors of A are at the same time the visitors of B.

To map the risk flows, we used the relative risk metric as explained in Tatem et al (2014): "For returning residents, it was assumed that the risk of acquiring an infection at their place of visit is a function of the level of risk at the visited location and the length of stay [25,29]. Therefore, a simple metric of cumulative risk was calculated by scaling the number of days spent at the visited location … by the modelled risk value there for each returning resident trip. For visitors to new locations … the relative risk of each visitor carrying an infection can be quantified by the estimated level of risk at their home locations. These simple metrics defined importation risk flow networks for returning residents, visitors and, by combining the two, overall risk flow…"

As the ultimate goal is identification of areas for targeted interventions, we also calculated the target effectiveness score as a sum of the decreases in import scores across the country that would appear if malaria incidence at the selected settlement would be brought to zero.

## 3. THE TOOL

Following the explained methodology, we developed a platform that transforms the daily updated CDR data into standard aggregates, complemented by malaria incidence data. These aggregates are fed into the interactive tool which can be accessed via an online interface.

The tool displays the current short-term mobility flows in the country, the latest reported incidence and the corresponding relative malaria export and import scores. For a preview, see Figure 1. Building on the risk scores, the target effectiveness score is provided for each area, complemented by a map of areas importing malaria from a selected area. In the last section, interactions between districts are displayed, enabling a quick overview of main exporters and importers per district. These insights enable the responsible district officials to understand better the malaria transmission dynamics in their country and coordinate their elimination efforts with other districts.

The tool is currently being piloted with the malaria district officials in the Southern Province of Zambia, in collaboration with the National Malaria Control Center and PATH, with the goal of scale up for the whole country before the next malaria season. The current set-up enables regular updates of the tool, showing always the most recent information for data-driven decisions.

Once tested in Zambia, the tool can be easily scalable to other countries, as telecom data is routinely collected by operators everywhere in a fairly standard format. Our platform can thus be plugged into any operator's system willing to contribute their data and, with only minor changes in the data cleaning scripts, can be quickly deployed and provide insights for any country.

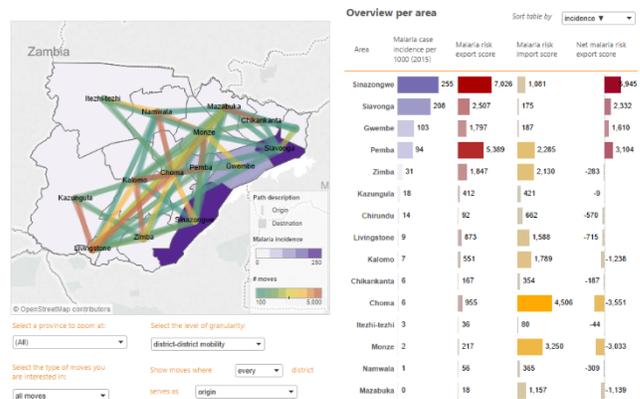

**Figure 1**: Malaria incidence and connectivity between different areas.

## 4. KNOWN CHALLENGES

There are several challenges connected to the work with CDRs, most of them shared by all studies using this data source. They include:

**Privacy.** Individual CDRs contain sensitive private information such as user's locations over time or their contacts to other people. In working with this data we stick to strict industry guidelines, as recommended by the GSMA (2014), where the raw individual's data never leaves the operator's premises and cannot be shared with external parties. Only geographically aggregated data is displayed in our tool, not enabling to single out any individuals and thus not compromising their privacy.

**Ownership bias.** Mobile phones are predominantly owned by wealthy young males (Blumenstock and Eagle, 2010). However, all segments of population still seem to be present in the user base, allowing for corrections during extrapolation and using CDRS as a robust mobility measure (Wesolowski et al., 2013). Similar correction is planned for future versions of our tool.

**Market share bias.** Typically, data from one operator only in one country is available for analysis, which means biases can be introduced due to different marketing strategy. In our case, we use data from one of the main operators in a country where no clear market stratification seems to be present.

**Cross-border mobility.** CDRs by default show data for one country only, ignoring any cross-border mobility. We currently do not account for cross-border mobility in any way, though it is an important factor for malaria transmission. In the future, we can analyze the appearance of foreign numbers in the dataset or try to match datasets from the same operator across the borders. Both of the solutions have challenges of their own though: many people switch to a local sim upon entering a foreign country, thus not using their original phone number, which makes their place of origin hard to track.

**Area receptivity.** In our current model, we do not take into account the broader context of temperature suitability of an area for mosquitoes, thus ignoring the fact that some areas are more receptive to incoming infections than others. We plan to introduce the temperature suitability estimation to the model in future versions of our tool.

**Activity-based records.** By using CDRs we are limited to considering people when they use their phone only, which means we are not recording substantial proportion of moves, simply because people did not use their phone at the given moment of time. Nevertheless, comparisons to other ways of measuring mobility show that CDR-based indicators actually show more mobility than is captured in traditional travel surveys (Wesolowski et al., 2014).

## 5. CONCLUSION

Telecom data can provide insights regarding people's mobility, which is priceless for epidemiological models of disease spreading. Based on previous research, we have built an operational tool fed with telecom data which shows malaria risk flows in Zambia in near real time. It provides malaria district officials with prioritization of areas based on their elimination effectiveness scores, as well as with detailed numbers of relative malaria exports and imports between different regions. With this information, the officials can build more efficient interventions and better target their information campaigns. Such information is particularly relevant for countries like Zambia, which are getting close to malaria elimination and need to prevent its re-introduction into areas that are already malaria-free.

Thanks to the nature of the telecom data, which is routinely collected in a standard format by telecom operators across the world, the tool is easily scalable to new geographies.

## 6. ACKNOWLEDGEMENTS

We would like to thank to MACEPA/PATH and the National Malaria Control Center in Zambia for their support and collaboration throughout this project, as well as to the telecom operator for enabling this project by providing the telecom data access within their CSR activities.

# 7.REFERENCES